\documentclass[preprint,superscriptaddress,nofootinbib]{revtex4}
     
     \usepackage{amssymb,amsmath,graphicx,color}
     \usepackage[colorlinks,breaklinks]{hyperref}
     \usepackage{multirow,natbib,rotating}
     \begin{document}
     \title{The feasibility of single ${\Lambda}$ production \\
     via ${\ell}^{-}$ $+$ $p$ ${\to}$ ${\Lambda}$ $+$ ${\nu}_{\ell}$
     at $e^{+}e^{-}$ colliders}
     \author{Yueling Yang}
     \affiliation{School of Physics, Henan Normal University, Xinxiang 453007, China}
     \author{Peisheng Tian}
     \affiliation{School of Physics, Henan Normal University, Xinxiang 453007, China}
     \author{Shuangshi Fang}
     \affiliation{Institute of High Energy Physics, Chinese Academy of Sciences, Beijing 100049, China}
     \affiliation{University of Chinese Academy of Sciences, Beijing 100049, China}
     \affiliation{Center for High Energy Physics, Henan Academy of Sciences, Zhengzhou 450046, China}
     \author{Bingbing Yang}
     \affiliation{School of Physics, Henan Normal University, Xinxiang 453007, China}
     \author{Junfeng Sun}
     \affiliation{School of Physics, Henan Normal University, Xinxiang 453007, China}

     %%%%% ----------------------------------------
     \begin{abstract}
     We present a comprehensive investigation of single ${\Lambda}$
     hyperon production via the lepton-nucleon deep inelastic scattering
     (LNDIS) process, ${\ell}^{-}$ $+$ $p$ ${\to}$ ${\nu}_{\ell}$ $+$ ${\Lambda}$,
     in the experimental environment of electron-positron colliders.
     Our approach utilizes incident leptons originating from the decays
     of resonances (${J/\psi}$, ${\psi}(2S)$, ${\Upsilon}(1S)$, ${\Upsilon}(2S)$,
     and $Z^{0}$) produced in $e^{+}e^{-}$ collisions, which then
     scatter off stationary protons in the surrounding detector materials.
     The differential and total cross sections are calculated using
     baryonic transition form factors parameterized with the
     $z$-expansion scheme within both the quantum chromodynamics
     (QCD) sum rule and lattice
     QCD frameworks. Our results indicate that the cross section
     increases with center-of-mass energy and is highly sensitive
     to the choice of form factors, resulting in significant
     theoretical uncertainties.
     This study highlights the experimental challenges in observing
     the LNDIS process at $e^{+}e^{-}$ colliders and underscores
     the need for improved determination of baryonic form factors.
     It serves as a valuable reference for future experimental
     searches and suggests that an anomalous observation of single
     ${\Lambda}$ hyperon production at $e^{+}e^{-}$ colliders
     could indicate new physics.

     \href{https://doi.org/10.1140/epjc/s10052-026-15878-8}{Eur. Phys. J. C 86, 663 (2026)}
     \end{abstract}
     %%%%% ----------------------------------------
     \maketitle

     \section{Introduction}
     The lepton-nucleon deep inelastic scattering (LNDIS) process,
     \begin{equation}
    {\ell}^{-} \, + \, p \, {\to} \,  {\nu}_{\ell} \, + \, {\Lambda}
     \label{ln2vy},
     \end{equation}
     %%%%% ------------------------------------
     serves as an essential experimental tool for probing
     the internal structure of baryons and investigating both
     strong and weak interactions. This process also has
     a significant impact on the proton-to-neutron
     number density ratio for cosmological archaeology.
     From the perspective of elementary particles and the
     basic interactions, the underlying dynamic mechanism of
     the LNDIS process in Eq.(\ref{ln2vy}) is analogous to
     that of the hyperon semileptonic decays (HSD),
     %%%%% ------------------------------------
     \begin{equation}
    {\Lambda} \, {\to} \,  p \, + \, {\ell}^{-} \, + \, \bar{\nu}_{\ell}
     \label{y2nlv}.
     \end{equation}
     %%%%% ------------------------------------
     Both LNDIS and HSD processes are actually induced by
     flavor-changing charged weak currents. Together,
     they play an important complementary role in testing  the
     universality of the couplings between the lepton flavors
     and the charged gauge bosons of the standard model (SM),
     exploring the potential deviations from the left-handed currents,
     and extracting the quark-mixing Cabibbo–Kobayashi–Maskawa (CKM)
     matrix element $V_{us}$.
     Improving the precision of $V_{us}$ measurement is of
     considerable significance.
     Currently, a $2.5 \, {\sigma}$ experimental discrepancy exists with
     the unitarity requirement,
     ${\vert} \, V_{ud} \, {\vert}^{2}$ $+$
     ${\vert} \, V_{us} \, {\vert}^{2}$ $+$
     ${\vert} \, V_{ub} \, {\vert}^{2}$ $=$
     $0.9983(7)$ \cite{pdg2024},
      which might hint at a possible  effect of new physics.

     Experimentally, measuring HSD is challenging due to the
     short lifetime ${\tau}_{\Lambda}$  $=$ $26.32(20)$ ns \cite{pdg2024},
     the undetectable neutrino, and the difficulty in distinguishing
     the signal from dominant two-body hadronic backgrounds.
     The distributions of the energy and momentum of
     the final states ${\ell}$ and $p$ in HSD are continuous.
     In contrast, in the LNDIS process, the energy and momentum
     of the produced ${\Lambda}$
     at the center-of-mass (CM) frame are well defined.
     This distinctive feature greatly facilitates the identification
     of the LNDIS signal, even when one of the final states,
     the neutrino, is undetected.
     Furthermore, the occurrence probability of the LNDIS process
     is kinematically enhanced by phase space factors.
     Specifically, while the phase space volume for HSD is limited by
     $m_{\Lambda}$ $-$ $m_{p}$ ${\approx}$ $180$ MeV, the phase
     space for the LNDIS process increases with higher CM energy.

\begin{figure}[h]
     \includegraphics[width=0.35\textwidth,bb=195 595 390 725]{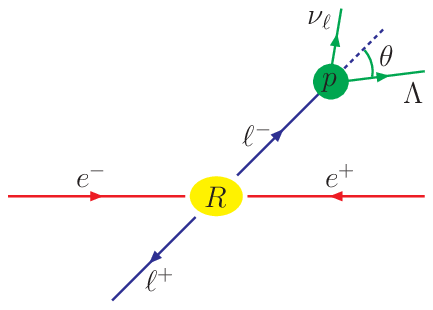}
     \caption{Diagrammatic sketch of the process
       ${\ell}^{-}$ $+$ $p$ ${\to}$ ${\Lambda}$ $+$ ${\nu}_{\ell}$
       at the $e^{+} \, e^{-}$ colliders.}
     \label{fig01}
     \end{figure}

     In this paper, we  focus on the single ${\Lambda}$
     production via the LNDIS process at $e^{+}e^{-}$ colliders.
     The proton in Eq.(\ref{ln2vy}) could be a material
     component of either the beam pipe or detectors
     surrounding the collision point.
     The initial charged lepton ${\ell}$ in Eq.(\ref{ln2vy}) with
     a specific energy can originate from both
     $e^{+} \, e^{-} \, {\to} \, {\ell}^{+} \, {\ell}^{-}$ and
     $e^{+} \, e^{-} \, {\to} \, R \, {\to} \, {\ell}^{+} \, {\ell}^{-}$,
     where, as illustrated in Fig. \ref{fig01},
     $R$ denotes resonances such as
     ortho-charmonium ${\psi}$, ortho-bottomonium ${\Upsilon}$ or
     gauge boson $Z^{0}$.
     The identification of the hyperon ${\Lambda}$
     together with the charged lepton ${\ell}^{+}$ on the opposite side of
     the incident ${\ell}^{-}$ can enhance both signal
     reconstruction efficiency and background suppression.
     The cross section ${\sigma}(e^{+} \, e^{-} \, {\to} \, {\ell}^{+} \, {\ell}^{-})$
     decreases with the increase in the $e^{+}e^{-}$ CM energy,
     except at the resonance peaks.
     The disintegration branching ratios of the ${\psi}$, ${\Upsilon}$,
     $Z^{0}$ ${\to}$ ${\ell}^{+} \, {\ell}^{-}$ decays as well as
     the expected yields of incident  ${\ell}$ available in the existing
     and future high-luminosity $e^{+} \, e^{-}$ collision experiments
     are listed in Tables \ref{br-psi-upsilon-z} and
     \ref{num-psi-upsilon-z}, respectively.
     Clearly, there is great potential and opportunity for
     investigating the LNDIS process at the $e^{+} \, e^{-}$ colliders.
     %%%%% ------------------------------------
     %%%%% ------------------------------------
     %%%%% ------------------------------------
     \begin{table}[h]
     \caption{The mass, width, and dilepton branching ratios ${\cal B}$
              of the particles of
              ${\psi}$, ${\Upsilon}$, $Z^{0}$ \cite{pdg2024}.}
     \label{br-psi-upsilon-z}
     \begin{ruledtabular}
     \begin{tabular}{c cc cc c}
       Resonance  & $J/{\psi}$ & ${\psi}(2S)$ & ${\Upsilon}(1S)$
                  & ${\Upsilon}(2S)$ & $Z^{0}$ \\ \hline
       Mass (GeV) & $3.096900(6)$ & $3.686097(10)$ & $9.46040(10)$
                  & $10.0234(5)$ & $91.1880(20)$ \\
       Width      & $92.6(1.7)$ keV
                  & $286(16)$ keV
                  & $54.02(1.25)$ keV
                  & $31.98(2.63)$ keV
                  & $2.4955(23)$  GeV  \\
       ${\cal B}( e^{+} \, e^{-} ) \, {\times} \, 10^{2} $
     & $5.971(32)$ & $0.794(22)$ & $2.39(8)$ & $1.91(16)$ & $3.3632(42)$ \\
       ${\cal B}( {\mu}^{+} \, {\mu}^{-} ) \, {\times} \, 10^{2} $
     & $5.961(33)$ & $0.80 (6 )$ & $2.48(4)$ & $1.93(17)$ & $3.3662(66)$
     \end{tabular}
     \end{ruledtabular}
     \end{table}
     %%%%% ------------------------------------
     %%%%% ------------------------------------
     \begin{table}[h]
     \caption{The expected numbers of the resonance $R$
              ${\in}$ $\{ {\psi}, \, {\Upsilon}, \, Z^{0} \}$ and
              the lepton ${\ell}^{\pm}$ from the $R$ decays
              available at the existing and future $ e^{+} \, e^{-} $
              collision experiments, where the number of
              $N_{ R \, {\to} \, {\ell}^{+} \, {\ell}^{-} }$ $=$
              $N_{R} \, {\times} \, {\cal B}( R \, {\to} \, {\ell}^{+} \, {\ell}^{-} )$
              with the branching ratio
              ${\cal B}( R \, {\to} \, {\ell}^{+} \, {\ell}^{-} )$
              given in Table \ref{br-psi-upsilon-z},
              the estimated STCF numbers are based on a data sample
              of $10 \, {\rm ab}^{-1}$ integrated luminosity;
              the Belle II numbers are estimated with $50$
              times the amount of the Belle data sample.}
     \label{num-psi-upsilon-z}
     \begin{ruledtabular}
     \begin{tabular}{c ccc}
        exp.
     & $ N_{ {\psi}(1S) } $ & $ N_{ {\psi}(1S) \, {\to} \, e^{+}     \, e^{-}     } $
                            & $ N_{ {\psi}(1S) \, {\to} \, {\mu}^{+} \, {\mu}^{-} } $
     \\ % \hline
        BESIII & $ 10.087(44) \, {\times} \, 10^{9} $ \cite{CPC.46.074001}
               & $  6.02(4) \, {\times} \, 10^{8} $
               & $  6.01(4) \, {\times} \, 10^{8} $ \\
        STCF   & $  3.4 \, {\times} \, 10^{13} $ \cite{stcf}
               & $  2.03(1) \, {\times} \, 10^{12} $
               & $  2.03(1) \, {\times} \, 10^{12} $
     \\ \hline
       exp.
     & $ N_{ {\psi}(2S) } $ & $ N_{ {\psi}(2S) \, {\to} \, e^{+}     \, e^{-}     } $
                            & $ N_{ {\psi}(2S) \, {\to} \, {\mu}^{+} \, {\mu}^{-} } $
     \\ % \hline
       BESIII  & $ 2.7124(143) \, {\times} \, 10^{9} $ \cite{CPC.48.093001}
               & $  2.15(06) \, {\times} \, 10^{7} $
               & $  2.17(16) \, {\times} \, 10^{7} $ \\
       STCF    & $  6.4 \, {\times} \, 10^{12} $ \cite{stcf}
               & $  5.08(14) \, {\times} \, 10^{10} $
               & $  5.12(38) \, {\times} \, 10^{10} $
     \\ \hline
       exp.
     & $ N_{ {\Upsilon}(1S) } $ & $ N_{ {\Upsilon}(1S) \, {\to} \, e^{+}     \, e^{-}     } $
                                & $ N_{ {\Upsilon}(1S) \, {\to} \, {\mu}^{+} \, {\mu}^{-} } $
     \\ % \hline
       Belle    & $ 1.02(2) \, {\times} \, 10^{8} $ \cite{EPJC.74.3026}
                & $ 2.44(9) \, {\times} \, 10^{6} $
                & $ 2.53(6) \, {\times} \, 10^{6} $ \\
       Belle II & $ 5.0 \, {\times} \, 10^{9} $     \cite{PTEP.2019.123c01}
                & $ 1.20(4) \, {\times} \, 10^{8} $
                & $ 1.24(2) \, {\times} \, 10^{8} $
     \\ \hline
       exp.
     & $ N_{ {\Upsilon}(2S) } $ & $ N_{ {\Upsilon}(2S) \, {\to} \, e^{+}     \, e^{-}     } $
                                & $ N_{ {\Upsilon}(2S) \, {\to} \, {\mu}^{+} \, {\mu}^{-} } $
     \\ % \hline
       Belle    & $ 1.58(4) \, {\times} \, 10^{8} $ \cite{EPJC.74.3026}
                & $ 3.02(26) \, {\times} \, 10^{6} $
                & $ 3.05(28) \, {\times} \, 10^{6} $ \\
       Belle II & $ 7.5 \, {\times} \, 10^{9} $
                & $ 1.43(12) \, {\times} \, 10^{8} $
                & $ 1.45(13) \, {\times} \, 10^{8} $
     \\ \hline
       exp.
     & $ N_{ Z^{0} } $ & $ N_{ Z^{0} \, {\to} \, e^{+}     \, e^{-}     } $
                       & $ N_{ Z^{0} \, {\to} \, {\mu}^{+} \, {\mu}^{-} } $
     \\ % \hline
       CEPC     & $ 2.5 \, {\times} \, 10^{12} $ \cite{cepc}
                & $ 8.41(1) \, {\times} \, 10^{10} $
                & $ 8.42(2) \, {\times} \, 10^{10} $ \\
       FCC-ee   & $ 6.0 \, {\times} \, 10^{12} $ \cite{fcc}
                & $ 20.18(3) \, {\times} \, 10^{10} $
                & $ 20.20(4) \, {\times} \, 10^{10} $
     \end{tabular}
     \end{ruledtabular}
     \end{table}
     %%%%% ------------------------------------

     \section{Cross section of LNDIS process}
     The differential cross section of the
     LNDIS process is given by,
     %%%%% ------------------------------------
     \begin{eqnarray}
     \frac{ d \,{\sigma} }{ d \, {\cos}{\theta} }  & = &
     \frac{1}{ 32 \, {\pi} } \,
     \frac{ k_{\rm c.m.} }{ \sqrt{s} } \,
     \frac{ {\vert} {\cal A}( {\ell}^{-} \, + \, p \, {\to} \,  {\nu}_{\ell} \, + \, {\Lambda} ) {\vert}^{2} }
          { \sqrt{ (p_{\ell}{\cdot}p_{p})^{2} - m_{\ell}^{2} \, m_{p}^{2} } }
     \nonumber \\ & = &
     \frac{ G_{F}^{2} \, {\vert} V_{us} {\vert}^{2} }{ 16 \, {\pi} \, s } \,
     \frac{ k_{\rm c.m.} }{ p_{\rm c.m.} } \,
     ( 1 - \frac{ m_{\ell}^{2} }{ q^{2} } ) %  \,  d \, {\cos}{\theta}
     \nonumber \\ & {\times} &
        \big\{ m_{\ell}^{2} \,
        \big[ H_{t,t} + ( H_{t,0} + H_{0,t} ) \, {\cos}{\theta}
     +  H_{0,0} \, {\cos}^{2}{\theta}
     +  \frac{ H_{+,+} + H_{-,-} }{2} \, {\sin}^{2}{\theta} \big]
     \nonumber \\ & + &  q^{2} \,
        \big[  H_{0,0} \, {\sin}^{2}{\theta}
            +  \frac{ H_{+,+} + H_{-,-} }{2} \, ( 1 + {\cos}^{2}{\theta} )
            -  ( H_{+,+} - H_{-,-} ) \, {\cos}{\theta} \big] \big\}
     \label{differential-cross-section-01},
     \end{eqnarray}
     %%%%% ------------------------------------
     with the square of the CM energy
     %%%%% ------------------------------------
     \begin{equation}
     s \, = \, (p_{\ell} + p_{p})^{2}
       \, = \, ( p_{\Lambda} + p_{\nu})^{2}
       \, = \, m_{R} \, m_{p} + m_{\ell}^{2} + m_{p}^{2}
     \label{cm-energy},
     \end{equation}
     %%%%% ------------------------------------
     and the CM momentum
     %%%%% ------------------------------------
     \begin{equation}
     p_{\rm c.m.} \, = \,
    {\vert} \vec{p}_{\ell}^{~\rm c.m.} {\vert} \, = \,
    {\vert} \vec{p}_{p}^{~\rm c.m.} {\vert} \, = \,
     \frac{ {\lambda}^{1/2}(s,m_{\ell}^{2},m_{p}^{2}) }{ 2 \, \sqrt{s} }
     \label{cm-momentum-u},
     \end{equation}
     %%%%% ------------------------------------
     %%%%% ------------------------------------
     \begin{equation}
     k_{\rm c.m.} \, = \,
    {\vert} \vec{p}_{\Lambda}^{~\rm c.m.} {\vert} \, = \,
    {\vert} \vec{p}_{\nu}^{~\rm c.m.} {\vert} \, = \,
     \frac{ {\lambda}^{1/2}(s,m_{\Lambda}^{2},0) }{ 2 \, \sqrt{s} }
     \label{cm-momentum-v},
     \end{equation}
     %%%%% ------------------------------------
     where ${\theta}$ is the angle between the momentum vector of
     the incident particle ${\ell}$ and the hyperon
     ${\Lambda}$, as illustrated in Fig.~\ref{fig01};
     ${\lambda}(a,b,c)$ $=$ $a^{2}$ $+$ $b^{2}$ $+$ $c^{2}$ $-$
     $2ab$ $-$ $2ac$ $-$ $2bc$ is the K\"{a}llen function;
     the four-momentum transfer $q$ $=$ $p_{p}$ $-$ $p_{\Lambda}$.
     We find that the widths of ${\psi}(1S,2S)$ and
     ${\Upsilon}(1S,2S)$ are too small relative to
     their mass to have perceivable effects on cross sections.
     The analytic expressions of scattering amplitude ${\cal A}$
     and helicity amplitude tensor $H_{{\lambda},{\lambda}^{\prime}}$
     are provided in Appendices \ref{app:amplitude} and \ref{app:hij-tensor},
     respectively.

     %%%%% ------------------------------------
     \begin{table}[h]
     \caption{Cross section of the process ${\ell}^{-}$ $+$ $p$ ${\to}$
          ${\Lambda}$ $+$ ${\nu}_{\ell}$ with the lepton ${\ell}^{-}$
          coming from different resonances,
          where the theoretical uncertainties come from
          the baryonic form factors, and
          the third uncertainties of the last column from
          the width of $Z^{0}$ boson, $m_{R}$ ${\in}$
          ($m_{Z^{0}}$ $-$ ${\Gamma}_{Z^{0}}/2$,
           $m_{Z^{0}}$ $+$ ${\Gamma}_{Z^{0}}/2$).}
     \label{tab:cs}
     \begin{ruledtabular}
     \begin{tabular}{ccc cc cc c}
      case & unit & process
     &  $J/{\psi}$ & ${\psi}(2S)$ & ${\Upsilon}(1S)$
     &  ${\Upsilon}(2S)$ & $Z^{0}$ \\ \hline
        \multirow{2}{*}{QCDSR}
      & \multirow{2}{*}{fb}
      & $e^{-}p$ ${\to}$ ${\Lambda}{\nu}$
      & $2.429^{+0.040}_{-0.032}$
      & $4.094^{+0.022}_{-0.015}$
      & $33.13 {\pm} 0.30 $
      & $36.52 {\pm} 0.33 $
      & $342.8 {\pm} 1.7 {\pm} 2.7 $ \\
      & & ${\mu}^{-}p$ ${\to}$ ${\Lambda}{\nu}$
      & $2.404^{+0.039}_{-0.032}$
      & $4.064^{+0.022}_{-0.014}$
      & $33.09^{+0.31}_{-0.30}$
      & $36.48^{+0.34}_{-0.33}$
      & $342.8 {\pm} 1.7 {\pm} 2.7 $ \\ \hline
        \multirow{2}{*}{LQCD}
      & \multirow{2}{*}{ab}
      & $e^{-}p$ ${\to}$ ${\Lambda}{\nu}$
      & $0.00^{+0.49}_{-0.00}$
      & $0.15^{+2.12}_{-0.15}$
      & $29.7^{+99.5}_{-29.7}$
      & $34.5^{+116.2}_{-~34.5}$
      & $1284^{+7269}_{-~766} {\pm} 23$ \\
      & & ${\mu}^{-}p$ ${\to}$ ${\Lambda}{\nu}$
      & $0.00^{+0.23}_{-0.00}$
      & $0.11^{+1.58}_{-0.11}$
      & $29.6^{+94.5}_{-29.6}$
      & $34.4^{+110.7}_{-~34.4}$
      & $1284^{+7242}_{-~802} {\pm} 23$
     \end{tabular}
     \end{ruledtabular}
     \end{table}
     %%%%% ------------------------------------
     %%%%% ------------------------------------
     \begin{figure}[h]
     \includegraphics[width=0.4\textwidth]{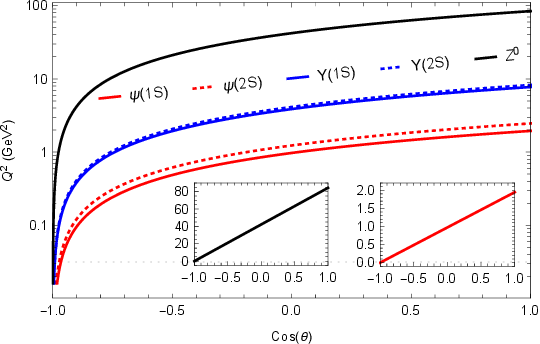}
     \caption{The distribution of $Q^{2}$ versus ${\cos}{\theta}$;
         the lines correspond to the different resonances $R$.}
     \label{fig:cos-q2}
     \end{figure}
     %%%%% ------------------------------------

     The numerical results of the cross section for the process
     ${\ell}^{-}$ $+$ $p$ ${\to}$ ${\Lambda}$ $+$ ${\nu}_{\ell}$
     at the $e^{+}e^{-}$ colliders are summarized in Table \ref{tab:cs}.
     Several remarks regarding these results are provided below.

     From Table~\ref{tab:cs},  it is clear that the cross section
     increases with CM energy, reaching its maximum when the
     incident ${\ell}^{-}$ originates from $Z^{0}$ boson decays.
     Additionally,  the cross section is highly sensitive to the
     values of the form factors. The cross sections calculated
     using the QCD sum rules (QCDSR) form factors \cite{PRD.112.094035}
     are about three to four orders of magnitude larger than those
     using lattice QCD (LQCD) form factors \cite{PRL.135.231901},
     even though both approaches can reasonably describe the
     experimental branching ratios for the HSD process ${\Lambda}$ ${\to}$
     $p\,{\ell}\,\bar{\nu}_{\ell}$ (${\ell}$ $=$ $e$ and ${\mu}$)
     within uncertainties \cite{PRD.112.094035,PRL.135.231901}.
     The baryonic form factors are scalar functions of $q^{2}$.
     For the HSD process, the physical region of time-like $q^{2}$
     is near zero, with
     $q^{2}_{\rm max}$ $=$ $(m_{\Lambda}-m_{p})^{2}$ ${\approx}$
     $0.03$ ${\rm GeV}^{2}$.
     In contrast, for the LNDIS process, the space-like
     $Q^{2}$ $=$ $-q^{2}$,  as a linear function of ${\cos}{\theta}$,
     spans a wide range,
    for example, $Q^{2}$ ${\in}$ $({\sim}0,{\sim}1.95)\,{\rm GeV}^{2}$
    [or $({\sim}0,{\sim}84.33)\,{\rm GeV}^{2}$] for ${\ell}^{-}$ from
     $J/{\psi}$ [or $Z^{0}$] decays,
     as illustrated in Fig. \ref{fig:cos-q2}.
     The $z$-expansion parameterization is
     usually employed to extrapolate form factors from the
     space-like $q^{2}$ region where no influences from
     any resonances appear, to the time-like $q^{2}$ domain.
     In this work, we employ the $z$-expansion
     parameterization \cite{PhysRevD.79.013008}
     for extrapolating the form factors from
     $q^{2}$ ${\approx}$ $0$ to a large $Q^{2}$ region,
     thereby avoiding unphysical behaviors associated with the
     polynomial parameterization.
     Because of $q^{2}$ ${\in}$ $[0,q^{2}_{\rm max}]$ for
     the HSD process, the $z$-expansion parameterization
     of ${\Lambda}$ ${\to}$ $p$ form factors,
     such as Refs. \cite{PRD.112.094035,PRL.135.231901,JHEP.2024.06.122},
     is calculated at a small interval, mainly concentrated in
     the $q^{2}$ ${\approx}$ $0$ vicinity.
     The data on the $q^{2}$ dependence of form factors
     in the $Q^{2}$ $>$ $0.1$ GeV$^{2}$ region are unavailable
     at this time. Therefore, it
     remains questionable whether extrapolating the form factors from
     $q^{2}$ ${\approx}$ $0$ to a large $Q^{2}$
     value is reliable, given that our current understanding
     of the form factors is based largely on their behavior near
     $q^{2}$ ${\approx}$ $0$. Therefore, precise measurements
     of the LNDIS process are essential to further constrain the
     form factors, which will aid in the determination of $V_{us}$.

     %%%%% ------------------------------------
     \begin{figure}[h]
     \includegraphics[width=0.4\textwidth]{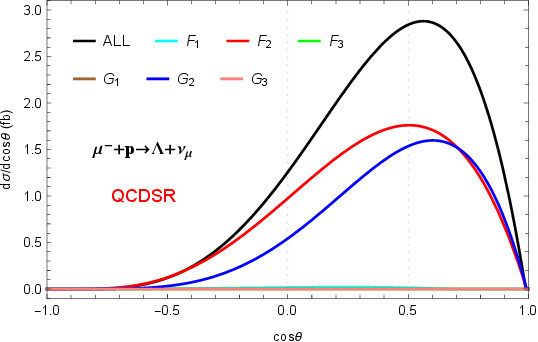}
     \caption{The distribution of $d{\sigma}/d{\cos}{\theta}$ versus
      ${\cos}{\theta}$, where the ``$F_{2}$'' (red) line denotes
      the contribution from only the form factor $F_{2}$ while
      the other form factors are set to zero,
      and the same applies to all other cases.
      The black line denotes the total contribution of all six
      from factors.}
     \label{fig:cos-cs}
     \end{figure}
     %%%%% ------------------------------------

    To better understand the contributions from different
    form factors, the distribution of the differential cross
    section $d{\sigma}/d{\cos}{\theta}$ is shown in
    Fig. \ref{fig:cos-cs}  using the central values of
    the QCDSR form factors \cite{PRD.112.094035} and
    for ${\mu}^{-}$ originating  from the purely leptonic
    $J/{\psi}$ decay. It is found that the main contributions
    to the cross section arise from the form factors $F_{2}$ and $G_{2}$,
    which is attributed to two aspects.  First, the magnitudes of
    $F_{2}$ and $G_{2}$ are relatively large in the $Q^{2}$ $>$ $1$
    GeV$^{2}$ region, as shown in Fig. \ref{fig:prd-ff}.
    Second, at large $Q^{2}$, the dominant terms in the cross
    section are proportional to $q^{2}$, while terms proportional
    to $m_{\ell}^{2}$ are suppressed (see Eq. (\ref{differential-cross-section-01})).
    Consequently, the main contributions stem from the hadronic
    helicity amplitude $H_{0,0}$, as illustrated in Fig. \ref{fig:hijq2}.
    In addition, the cross section is mainly concentrated in the
    forward angle region, which resembles the behavior of a
    $t$-channel process. However, the contribution of $q^{2}\, H_{0,0}$
    is suppressed by a factor of ${\sin}^{2}{\theta}$ as ${\theta}$ ${\to}$ $0$.

     \section{Expectations at the $e^+e^-$ colliders}
     The number of single ${\Lambda}$ hyperon events
     is given by the product of the cross section ${\sigma}$ and
     the time integrated luminosity $L$,
     %%%%% ------------------------------------
     \begin{equation}
     N_{\rm sig} \, = \, {\sigma} \, {\times} \, L
     \, = \, {\sigma} \, {\times} \, {\int} {\Phi} \, n_{t} \, dt
     \label{definition-signal},
     \end{equation}
     %%%%% ------------------------------------
     where ${\Phi}$ is the beam flux of incident particle ${\ell}^{-}$;
     and here ${\int} {\Phi} dt$ $=$ $N_{\ell}$ is the number of incident
     particles listed in Table \ref{num-psi-upsilon-z}.
     $n_{t}$ is the proton areal density of the target,
     %%%%% ------------------------------------
     \begin{equation}
     n_{t} \, = \, \sum\limits_{i} n_{i}
     \, = \, \sum\limits_{i} {\rho}_{i} \, d_{i} \, \, N_{A} \, Z_{i} / M_{i}
     \label{target-areal-density},
     \end{equation}
     %%%%% ------------------------------------
     where ${\rho}_{i}$ is the density,
     $d_{i}$ is the thickness,
     $N_{A}$ $=$ $6.022\,140\,76\,{\times}\,10^{23}\,{\rm mol}^{-1}$
     is the Avogadro constant,
     $Z_{i}$ is the atomic number,
     and $M_{i}$ is the molar mass.
     Following the scenarios described in Ref. \cite{PRL.130.251902},
     we takee the stationary protons in the beam pipe at the
     Beijing Spectrometer III (BESIII) experiment as an example and assume that
     the $e^{+}e^{-}$ collision vertex  is fully enclosed by the beam pipe.
     The schematic diagram of the beam pipe of the BESIII detector
     is shown in Fig. 1 of Ref. \cite{PRL.130.251902}, and the
     beam pipe is composed of gold (${}^{197}_{~79}$Au),
     beryllium (${}^{9}_{4}$Be), and oil (${}^{12}_{~6}$C:${}^{1}_{1}$H
     $=$ $1$:$2.13$) \cite{PRL.130.251902}.
     The effective proton target density of the BESIII beam pipe is
     estimated as follows\footnotemark[1],
     \footnotetext[1]{According to the data of Ref. \cite{element},
     the densities of elements (in units of ${\rm g}/{\rm cm}^{3}$) are
     ${\rho}_{\rm H}$ $=$ $8.988\,{\times}10^{-5}$,
     ${\rho}_{\rm Be}$ $=$ $1.85$,
     ${\rho}_{\rm C}$ $=$ $2.267$,
     ${\rho}_{\rm Au}$ $=$ $19.282$.
     According to the description of Ref. \cite{PRL.130.251902},
     the vertical thicknesses (in units of ${\rm cm}$) are
     $d_{\rm H}$ $=$ $0.08$,
     $d_{\rm Be}$ $=$ $0.14$,
     $d_{\rm C}$ $=$ $0.08$,
     $d_{\rm Au}$ $=$ $0.001436$.}
     %%%%% ------------------------------------
     \begin{equation}
     n_{t}
     \ {\approx} \ 4 \, {\pi} \, ( n_{\rm Au} + n_{\rm Be} + n_{\rm Oil} )
     \ {\approx} \ 1.64 \, {\rm b}^{-1}
     \label{target-number}.
     \end{equation}
     %%%%% ------------------------------------
     With the statistics of over $10^{10}$ $J/{\psi}$  events
     and the cross section calculated with the QCDSR approach in
     Table \ref{tab:cs}, the expected signal yield is approximately
     $2\,{\times}\,10^{-6}$.
     Assuming the areal density given by Eq.(\ref{target-number}) is
     also applicable for the future Super Tau Charm Facility (STCF),
     Circular Electron Positron Collider (CEPC),
     and Future Circular Collider (FCC-ee)
     experiments, and using the incident lepton numbers from
     Table \ref{num-psi-upsilon-z} together with the QCDSR cross section
     in Table \ref{tab:cs}, the expected numbers of events are
     about $0.01$, $0.05$ and $0.1$, respectively.
     In this case, the observation
     of ${\ell}^{-}$ $+$ $p$ ${\to}$ ${\nu}_{\ell}$ $+$ ${\Lambda}$
     at BESIII, Belle-II, STCF, CEPC and FCC-ee experiments
     could potentially indicate new physics. However, these
     results should be interpreted with caution,
     as significant uncertainties remain in the
     extrapolation of baryonic form factors.

     \section{Summary}
     This paper presents an investigation of
     single ${\Lambda}$ hyperon production via the process
     ${\ell}^{-}$ $+$ $p$ ${\to}$ ${\Lambda}$ $+$ ${\nu}_{\ell}$,
     with the unique experimental environment of electron-positron
     colliders. We propose a novel approach in which the incident
     leptons ($e^{-}$, ${\mu}^{-}$) are generated from the decays
     of resonances ($J/\psi$, $\psi(2S)$, $\Upsilon(1S)$,
     $\Upsilon(2S)$ and $Z^0$) produced in $e^{+}e^{-}$
     collisions, and then scatter off stationary
     protons in detector materials.

     When the incident particle ${\ell}^{-}$ is produced from
     the $Z^{0}$ boson decays, the cross section of the process
     ${\ell}^{-}$ $+$ $p$ ${\to}$ ${\Lambda}$ $+$ ${\nu}_{\ell}$
     is estimated to be  $340$ ($1$) fb with the  QCDSR (LQCD)
     baryonic form factors determined via the $z$-series
     extrapolation.

    Despite the substantial fluxes of incident leptons at current
    (BESIII, Belle II) and future high-luminosity facilities (STCF, CEPC, FCC-ee),
    our feasibility analysis suggests that the expected event
    rates are extremely low. Even with optimistic cross section
    estimates based on QCDSR form factors and after accounting
    for realistic  target density estimates from beam pipe materials,
    the predicted number of observable signal events remains well
    below $1$  across all examined experimental scenarios.

     In conclusion, the data on the $q^{2}$ dependence
     of form factors in the deep space-like region are critically needed
     but important for theoretical explanation of the HSD and
     LNDIS processes.  Although the process
     ${\ell}^{-}$ $+$ $p$ ${\to}$ ${\Lambda}$ $+$ ${\nu}_{\ell}$
     could provide valuable constraints on baryonic transition form
     factors and the CKM matrix element $V_{us}$,
     its experimental observation at $e^{+}e^{-}$ colliders
     appears highly challenging under current conditions.
     Our results underscore the need for more precise
     determinations of baryonic form factors and
     suggest that any future observation of this process
     would likely point to significant deviations from
     present theoretical predictions.

     \section*{Acknowledgments}
     This work is supported by the National Natural Science Foundation of China (Grant Nos. 12275068, 12225509),
     the National Key R\&D Program of China (Grant No. 2023YFA1606000),
     and the Natural Science Foundation of Henan Province (Grant Nos. 262300421354, 252300421491, 242300420250).

     \begin{appendix}
     %%%%% ----------------------------------------
     \section{the scattering amplitude}
     \label{app:amplitude}
     The scattering amplitude for Eq.(\ref{ln2vy})
     is expressed as \cite{zpc46.p93},
     %%%%% ----------------------------------------
     \begin{equation}
    {\cal A} \, = \,
    {\langle} \, {\Lambda} \, {\nu} \, {\vert} \, {\cal H}_{\rm eff} \,
    {\vert} \, {\ell} \, p \,  {\rangle}
     \, = \,
     \frac{ G_{F} }{ \sqrt{2} } \, V_{us}^{\ast} \,
     \sum\limits_{ {\lambda}, \, {\lambda}^{\prime} } \,
     L_{\lambda}(\vec{p}_{W}) \, H_{{\lambda}^{\prime}}(\vec{p}_{W}) \,
     g_{{\lambda},{\lambda}^{\prime}}
     \label{eq:amplitude-helicity}
     \end{equation}
     %------------------------------------
     where the parameter $G_{F}$ ${\approx}$
     $ 1.166 \, {\times} \, 10^{-5} \, {\rm GeV}^{-2} $ \cite{pdg2024}
     is the Fermi constant;
     the CKM element $V_{us}^{\ast}$ describes the strength
     of the $u$ ${\to}$ $s$ transition,
     ${\vert} V_{us} {\vert}$ $=$ $0.22431(85)$ \cite{pdg2024}.
     $L_{\lambda}$ and $H_{{\lambda}^{\prime}}$
     represent leptonic and hadronic helicity amplitudes,
     respectively, and defined as follows,
     %%%%% ----------------------------------------
     \begin{equation}
     L_{\lambda}(\vec{p}_{W})
     \, = \,
    {\varepsilon}_{W}^{{\alpha}}({\lambda},\vec{p}_{W}) \, L_{\alpha}
     \label{leptonic-helicity-amplitude},
     \end{equation}
     %%%%% ----------------------------------------
     \begin{equation}
     H_{{\lambda}^{\prime}}(\vec{p}_{W}) \, = \,
    {\varepsilon}_{W}^{{\ast}{\alpha}}({{\lambda}^{\prime}},\vec{p}_{W}) \, H_{\alpha}
     \label{hadronic-helicity-amplitude},
     \end{equation}
     %%%%% ----------------------------------------
     %%%%% ----------------------------------------
     \begin{equation}
     L_{\alpha} \, = \,
    {\langle} \, {\nu} \, {\vert} \, \bar{\nu} \, {\gamma}_{\alpha} \,
     ( 1 - {\gamma}_{5} ) \, {\ell} \, {\vert} \, {\ell} \, {\rangle}
     \label{leptonic-matrix},
     \end{equation}
     %%%%% ----------------------------------------
     \begin{equation}
     H_{\alpha} \, = \,
    {\langle} \, {\Lambda} \, {\vert} \, \bar{s} \, {\gamma}_{\alpha} \,
     ( 1 - {\gamma}_{5} ) \, u \, {\vert} \, p \, {\rangle}
     \label{hadronic-matrix},
     \end{equation}
     %%%%% ----------------------------------------
     where ${\varepsilon}_{W}^{{\alpha}}({\lambda},\vec{p}_{W})$
     denotes the polarization vectors of the virtual $W^{\ast}$
     boson; ${\lambda}$, ${\lambda}^{\prime}$ $=$ $t$, $+$, $-$, $0$
     correspond to the helicity components, and the metric tensor
     $g_{ {\lambda}, {\lambda}^{\prime} }$ $=$ ${\rm diag}(+1,-1,-1,-1)$.

     %%%%% ----------------------------------------
     \section{the baryonic form factors}
     \label{app:formfactor}
     Phenomenologically, the hadronic matrix of
     Eq.(\ref{hadronic-matrix}) is usually parameterized
     in terms of six baryonic transition form factors
     based on the Lorentz structure
     \cite{ARNPS.34.351,ARNPS.53.39,
     PR.120.1482,ZP.212.190,NPB.12.669,
     JHEP.2024.06.122},
     %%%%% ------------------------------------
     \begin{eqnarray}
     H_{\alpha} & = &
     \bar{u}_{\Lambda}(p_{\Lambda}) \, \big\{ \big[ \,
            f_{1}(q^{2}) \, {\gamma}_{\alpha}
     - i \, f_{2}(q^{2}) \, {\sigma}_{{\alpha}{\beta}} \, \frac{ q^{\beta} }{ m_{p} } \,
     +      f_{3}(q^{2}) \,  \frac{ q_{\alpha} }{ m_{p} }  \big]
     \nonumber \\ & & \qquad\, \ + \  \big[ \,
            g_{1}(q^{2}) \, {\gamma}_{\alpha}
     - i \, g_{2}(q^{2}) \, {\sigma}_{{\alpha}{\beta}} \, \frac{ q^{\beta} }{ m_{p} } \,
     +      g_{3}(q^{2}) \,  \frac{ q_{\alpha} }{ m_{p} }
     \big] \, {\gamma}_{5} \, \big\} \, u_{p}(p_{p})
     \label{formfactor-tensor},
     \end{eqnarray}
     %%%%% ------------------------------------
     or the equivalent form
     \cite{PR.120.1482,ZP.212.190,NPB.12.669},
     %%%%% ------------------------------------
     \begin{eqnarray}
     H_{\alpha} & = &
     \bar{u}_{\Lambda}(p_{\Lambda}) \, \big\{ \big[ \,
        F_{1}(q^{2}) \, {\gamma}_{\alpha}
     +  F_{2}(q^{2}) \   \frac{ p_{{p},{\alpha}} }{ m_{p} } \,
     +  F_{3}(q^{2}) \   \frac{ q_{\alpha} }{ m_{p} }  \big]
     \nonumber \\ & & \qquad\, \ + \  \big[ \,
        G_{1}(q^{2}) \, {\gamma}_{\alpha}
     +  G_{2}(q^{2}) \,  \frac{ p_{{p},{\alpha}} }{ m_{p} } \,
     +  G_{3}(q^{2}) \,  \frac{ q_{\alpha} }{ m_{p} }
     \big] \, {\gamma}_{5} \, \big\} \, u_{p}(p_{p})
     \label{formfactor-vector},
     \end{eqnarray}
     %%%%% ------------------------------------
     where $\bar{u}_{\Lambda}$ and $u_{p}$ are the Dirac spinors
     of the baryons ${\Lambda}$ and $p$, respectively.
     The relations between the two sets of form factors
     are \cite{PR.120.1482,ZP.212.190,NPB.12.669},
     %%%%% ------------------------------------
     \begin{equation}
     F_{1} \, = \, f_{1} + \frac{ m_{\Lambda} + m_{p} }{ m_{p} } \, f_{2},
     \qquad
     F_{2} \, = \, - 2 \, f_{2},
     \qquad
     F_{3} \, = \, f_{2} + f_{3},
     \label{formfactor-relation-F1}
     \end{equation}
     %%%%% ------------------------------------
     \begin{equation}
     G_{1} \, = \, g_{1} + \frac{ m_{\Lambda} - m_{p} }{ m_{p} } \, g_{2},
     \qquad
     G_{2} \, = \, - 2 \, g_{2},
     \qquad
     G_{3} \, = \, g_{2} + g_{3}.
     \label{formfactor-relation-G1}
     \end{equation}
     %%%%% ------------------------------------

     A comprehensive collection of these baryonic form factors
     can be found in Refs. \cite{JHEP.2024.06.122,PRL.135.231901}.
     As we know, all six form factors were given by
     QCD sum rules \cite{PRD.112.094035} and
     lattice QCD \cite{PRL.135.231901}.

     In Ref. \cite{PRD.112.094035},
     the baryonic form factors are defined as follows.
     %%%%% ----------------------------------------
     \begin{eqnarray}
    {\langle} \, p \, {\vert} \, V^{\mu} \, {\vert} \, {\Lambda} \, {\rangle}
     & = &
     \bar{u}_{p}(p_{p}) \, \big[ \,
       F_{1}^{a}(q^{2}) \, {\gamma}^{\mu}
     + F_{2}^{a}(q^{2}) \,  \frac{ p_{\Lambda}^{\mu} }{ m_{\Lambda} } \,
     + F_{3}^{a}(q^{2}) \,  \frac{ p_{p}^{\mu} }{ m_{p} } \, \big] \, u_{\Lambda}(p_{\Lambda})
     \label{prd-e4v}, \\
    {\langle} \, p \, {\vert} \, A^{\mu} \, {\vert} \, {\Lambda} \, {\rangle}
     & = &
     \bar{u}_{p}(p_{p}) \, \big[ \,
       G_{1}^{a}(q^{2}) \, {\gamma}^{\mu}
     + G_{2}^{a}(q^{2}) \,  \frac{ p_{\Lambda}^{\mu} }{ m_{\Lambda} } \,
     + G_{3}^{a}(q^{2}) \,  \frac{ p_{p}^{\mu} }{ m_{p} } \,
     \big] \, {\gamma}_{5} \, u_{\Lambda}(p_{\Lambda})
     \label{prd-e4a},
     \end{eqnarray}
     %%%%% ----------------------------------------

     The relations of form factors between Eq.(\ref{prd-e4v}),
     Eq.(\ref{prd-e4a}) and Eq.(\ref{formfactor-vector}) are
     as follows.
     %%%%% ------------------------------------
     \begin{equation}
     F_{1} \, = \, + F_{1}^{a},
     \qquad
     F_{2} \, = \, + \frac{ m_{p} }{ m_{\Lambda} } \, F_{2}^{a} + F_{3}^{a},
     \qquad
     F_{3} \, = \, - \frac{ m_{p} }{ m_{\Lambda} } \, F_{2}^{a},
     \label{formfactor-prd-F1}
     \end{equation}
     %%%%% ------------------------------------
     %%%%% ------------------------------------
     \begin{equation}
     G_{1} \, = \, - G_{1}^{a},
     \qquad
     G_{2} \, = \, + \frac{ m_{p} }{ m_{\Lambda} } \, G_{2}^{a} + G_{3}^{a},
     \qquad
     G_{3} \, = \, - \frac{ m_{p} }{ m_{\Lambda} } \, G_{2}^{a}.
     \label{formfactor-prd-G1}
     \end{equation}
     %%%%% ------------------------------------

\begin{figure}[h]
     \includegraphics[width=0.45\textwidth]{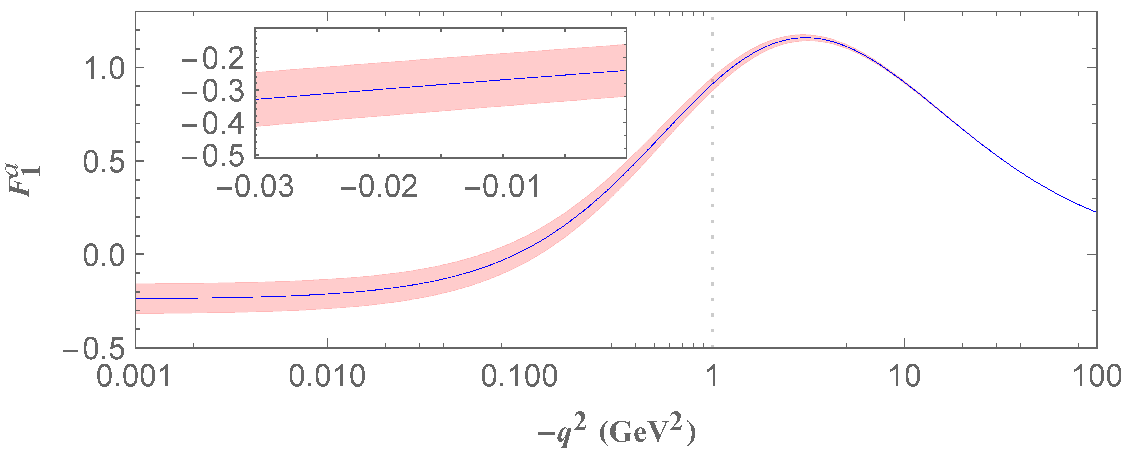} \ \ \
     \includegraphics[width=0.45\textwidth]{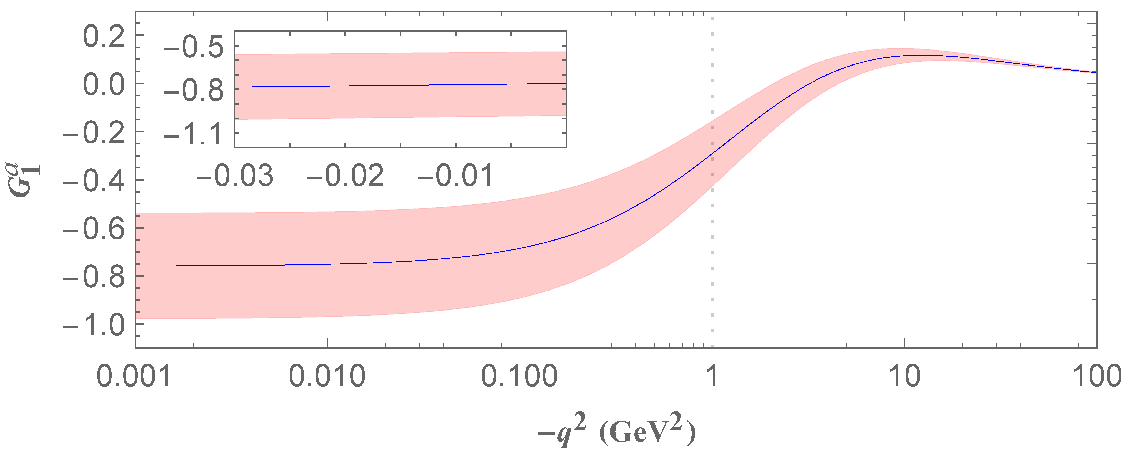} \\
     \includegraphics[width=0.45\textwidth]{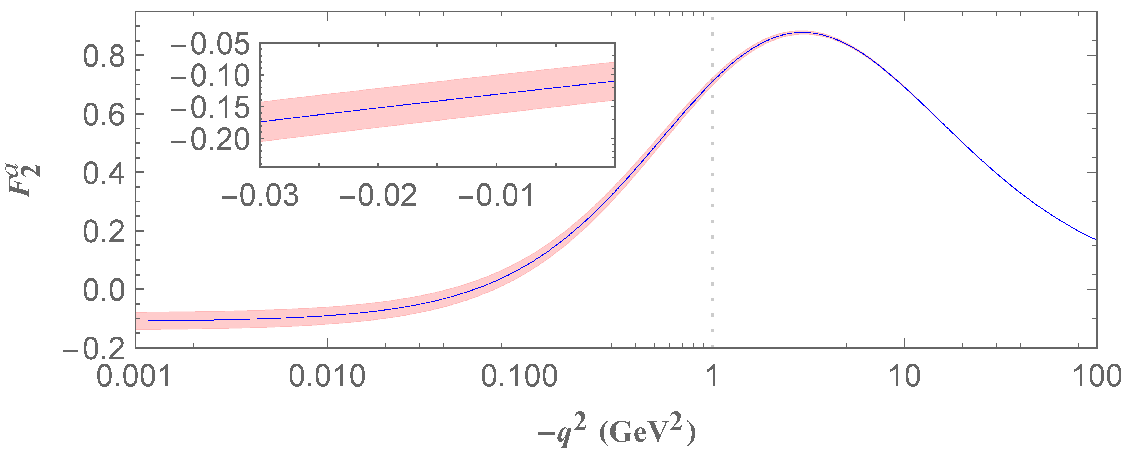} \ \ \
     \includegraphics[width=0.45\textwidth]{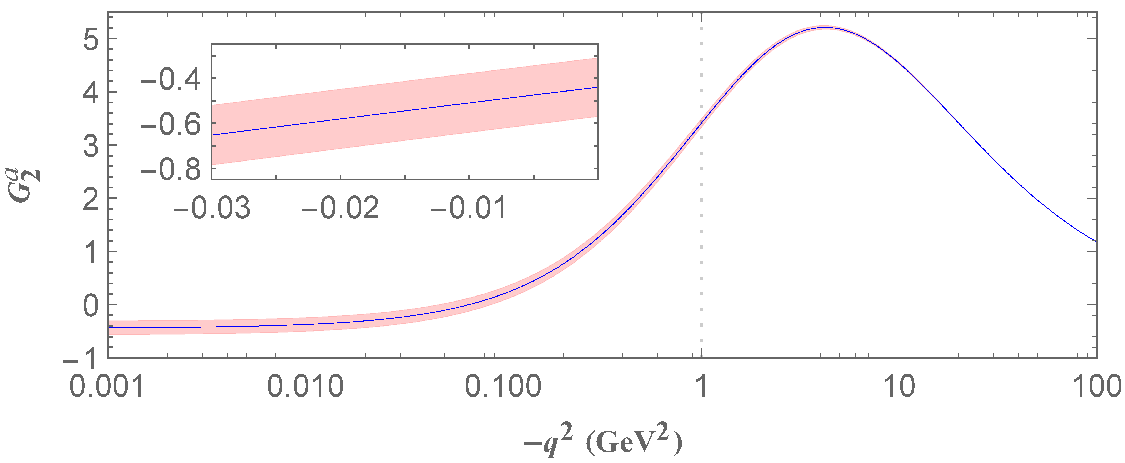} \\
     \includegraphics[width=0.45\textwidth]{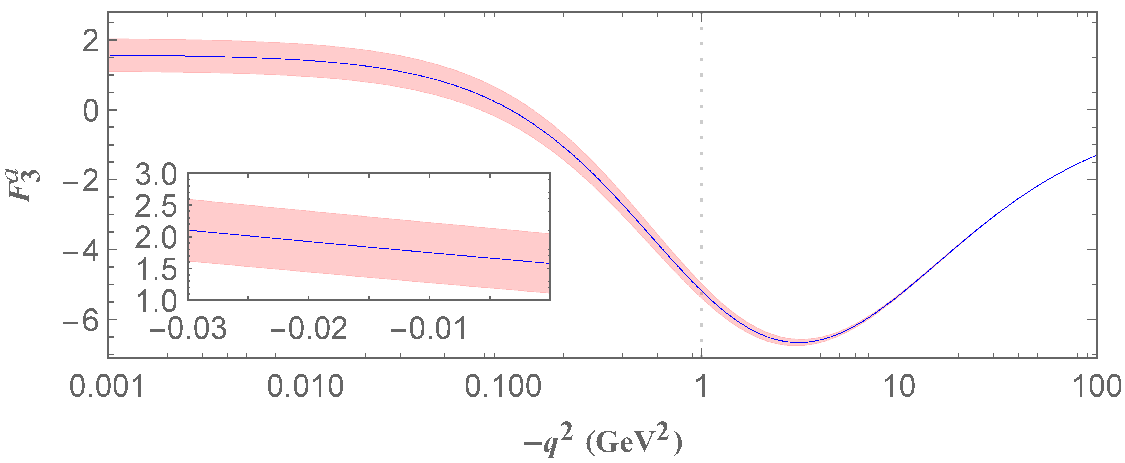} \ \ \
     \includegraphics[width=0.45\textwidth]{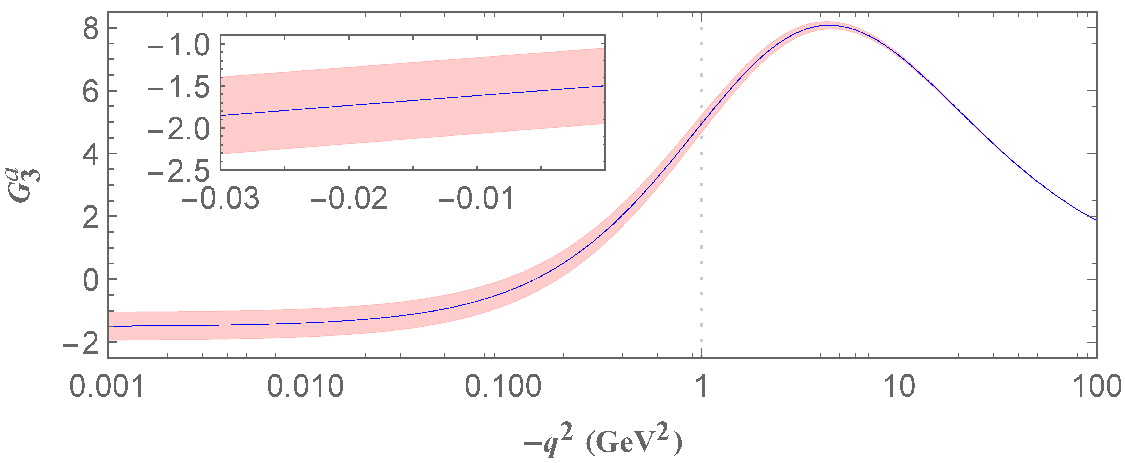}
     \caption{The shape line of the form factors $F_{i}^{a}$
       versus $q^{2}$, where the solid lines (bands)
       correspond to the center values (uncertainties)
       of $F_{i}^{a}$,
       and the embedded images are for making comparisons
       with FIG. 8 of Ref. \cite{PRD.112.094035}.}
     \label{fig:prd-ff}
     \end{figure}

     With  Eq.(27) and the $z$-expansion coefficients in
     TABLE III of  Ref. \cite{PRD.112.094035},
     the shape lines of form factors are shown in Fig. \ref{fig:prd-ff}.
     %%%%% ------------------------------------
     %%%%% ------------------------------------

\begin{figure}[h]
     \includegraphics[width=0.45\textwidth]{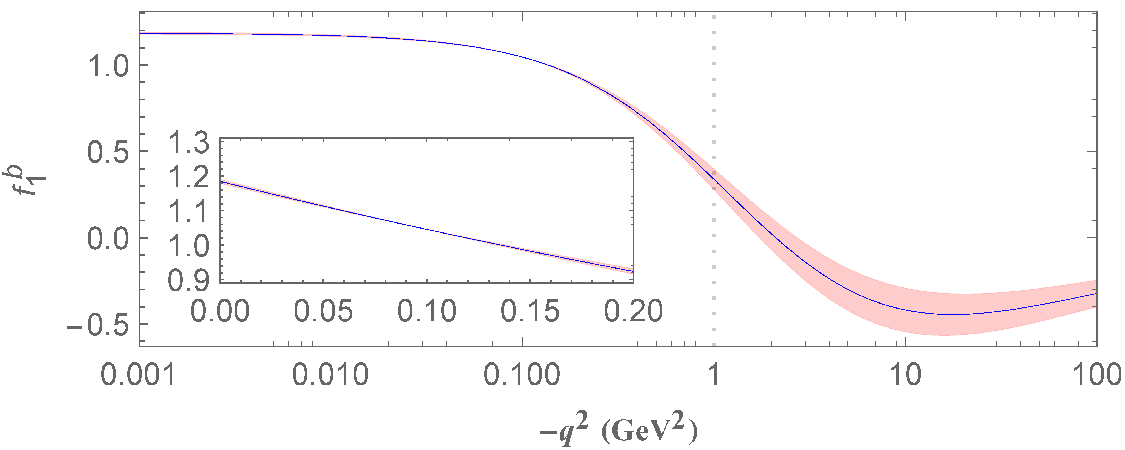} \ \ \
     \includegraphics[width=0.45\textwidth]{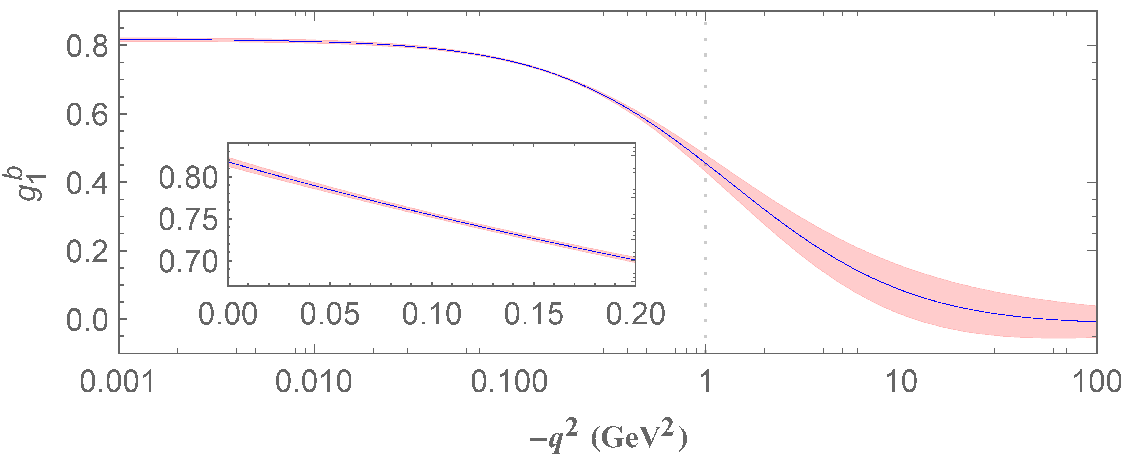} \\
     \includegraphics[width=0.45\textwidth]{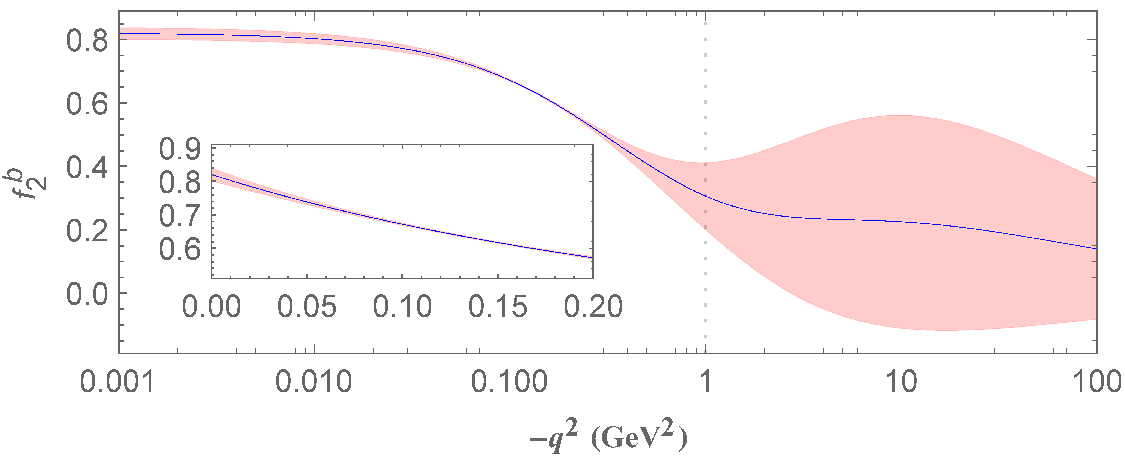} \ \ \
     \includegraphics[width=0.45\textwidth]{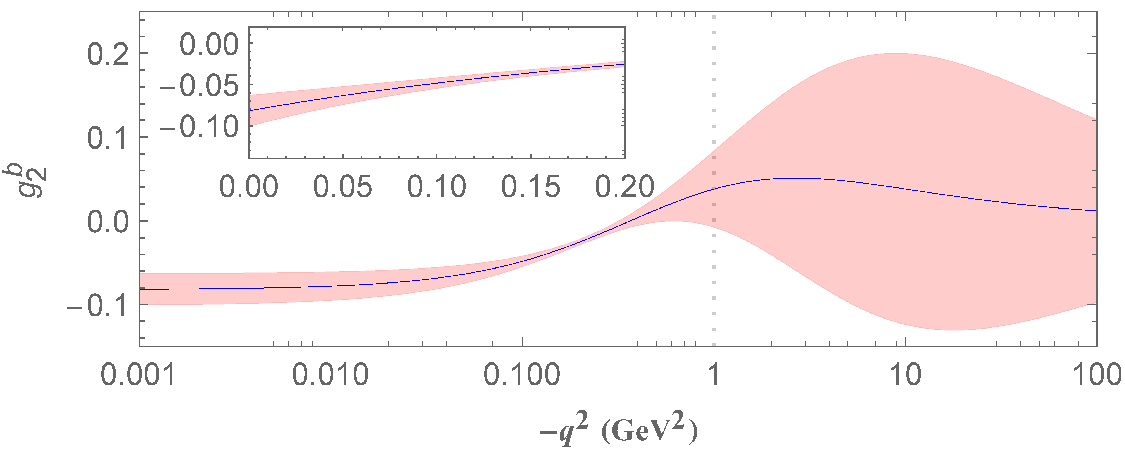} \\
     \includegraphics[width=0.45\textwidth]{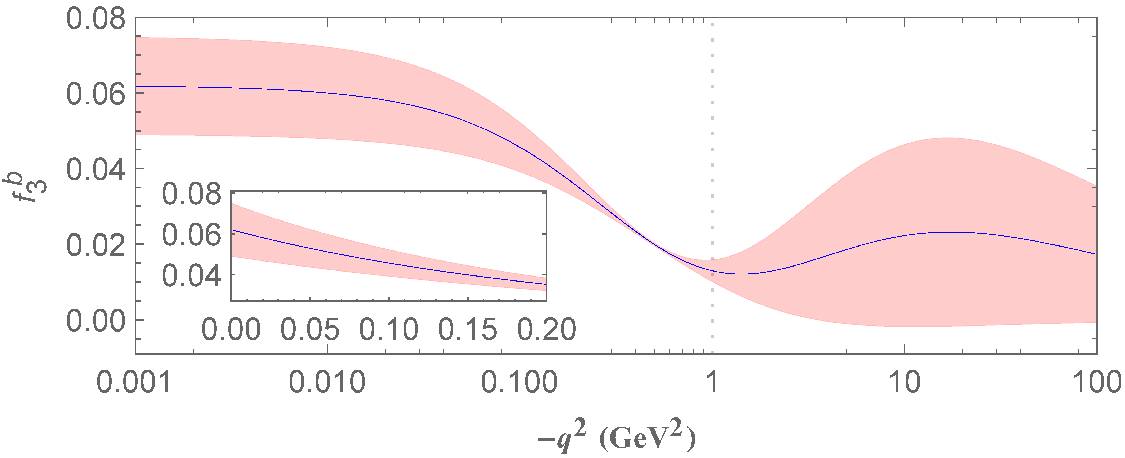} \ \ \
     \includegraphics[width=0.45\textwidth]{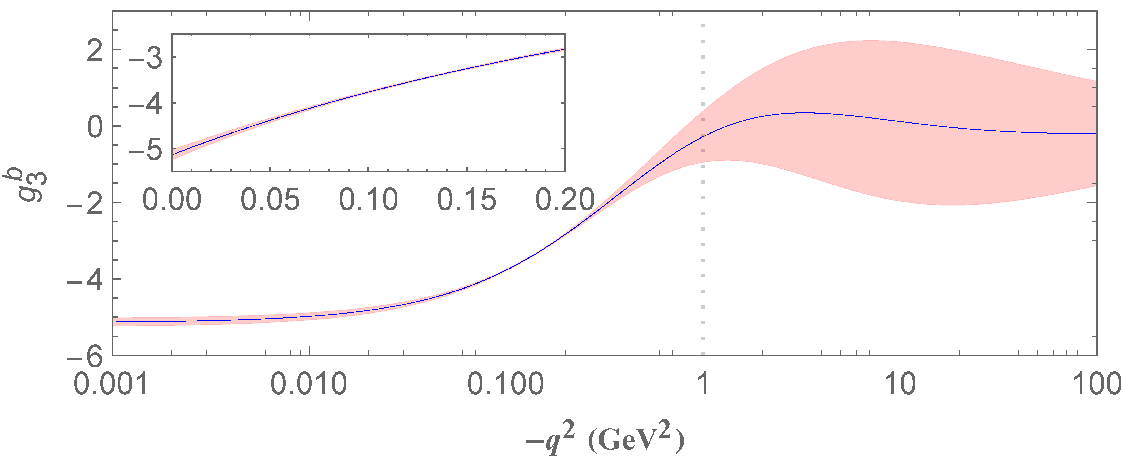}
     \caption{The shape line of the form factors $f_{i}^{b}$
       versus $q^{2}$, where the solid lines (bands)
       correspond to the center values (uncertainties)
       of the coefficients of the $z$-expansion,
       and the embedded images are for making comparisons
       with FIG. 1 of Ref. \cite{PRL.135.231901}.}
     \label{fig:prl-ff}
     \end{figure}

     In Ref. \cite{PRL.135.231901},
     the baryonic form factors are defined as follows.
     %%%%% ----------------------------------------
     \begin{eqnarray}
    {\langle} \, p \, {\vert} \, V_{\mu}-A_{\mu} \, {\vert} \, {\Lambda} \, {\rangle}
     & = &
     \bar{u}_{p}(p_{p}) \, \big\{ \big[ \,
       f_{1}^{b} \, {\gamma}_{\mu}
     - i \, \frac{ f_{2}^{b} }{ m_{\Lambda} } \, {\sigma}_{{\mu}{\nu}} \, q^{\nu}
     + \frac{ f_{3}^{b} }{ m_{\Lambda} } \, q_{\mu} \, \big] \, u_{\Lambda}(p_{\Lambda})
     \nonumber  \\ &  & \qquad \ - \, \big[ \,
              g_{1}^{b} \, {\gamma}_{\mu}
     - i \, \frac{ g_{2}^{b} }{ m_{\Lambda} } \, {\sigma}_{{\mu}{\nu}} \, q^{\nu}
     + \frac{ g_{3}^{b} }{ m_{\Lambda} } \, q_{\mu} \,
     \big] \, {\gamma}_{5} \, \big\} \, u_{\Lambda}(p_{\Lambda})
     \label{prl-e7}.
     \end{eqnarray}
     %%%%% ----------------------------------------
     The relations of form factors between Eq.(\ref{prl-e7})
     and Eq.(\ref{formfactor-tensor}) are as follows.
     %%%%% ------------------------------------
     \begin{equation}
     f_{1} \, = \, + f_{1}^{b},
     \qquad
     f_{2} \, = \, + \frac{ m_{p} }{ m_{\Lambda} } \, f_{2}^{b},
     \qquad
     f_{3} \, = \, - \frac{ m_{p} }{ m_{\Lambda} } \, f_{3}^{b},
     \label{formfactor-prl-f1}
     \end{equation}
     %%%%% ------------------------------------
     %%%%% ------------------------------------
     \begin{equation}
     g_{1} \, = \, - g_{1}^{b},
     \qquad
     g_{2} \, = \, + \frac{ m_{p} }{ m_{\Lambda} } \, g_{2}^{b},
     \qquad
     g_{3} \, = \, - \frac{ m_{p} }{ m_{\Lambda} } \, g_{3}^{b}.
     \label{formfactor-prl-g1}
     \end{equation}
     %%%%% ------------------------------------
     The shape lines of form factors of  Ref. \cite{PRL.135.231901}
     are shown in Fig. \ref{fig:prl-ff}.
     %%%%% ------------------------------------
     %%%%% ------------------------------------

     %%  The asymptotic behavior of form factors in Fig. \ref{fig:prd-ff}
     %%  and \ref{fig:prl-ff} are basically in accord with the customary
     %%  to constrain the $z$-expansion to vanish in the large
     %%  $Q^{2}$ $=$ $-q^{2}$ \cite{PhysRevD.79.013008},
     %%  although the control of the LQCD uncertainties in the
     %%  large $Q^{2}$ regions still need improving.

     %%%%% ----------------------------------------
     \section{the baryonic helicity amplitude tensor}
     \label{app:hij-tensor}
     %%%%% ------------------------------------
     The baryonic helicity amplitude tensors are written as,
     %%%%% ------------------------------------
     \begin{equation}
     H_{{\lambda},{\lambda}^{\prime}} \, = \, \frac{1}{4} \, \sum\limits_{s_{p}} \,
     \sum\limits_{s_{\Lambda}} \, H_{\lambda} \, H^{\dagger}_{{\lambda}^{\prime}}
     \label{hadronic-helicity-tensor},
     \end{equation}
     %%%%% ------------------------------------
     where the helicity amplitude $H_{\lambda}$ is defined
     by Eq.(\ref{hadronic-helicity-amplitude}).

     %%%%% ------------------------------------
     \begin{eqnarray}
     H_{+,+} & = &
       ( p_{p} \, {\cdot} \, p_{\Lambda} - m_{p} \, m_{\Lambda} ) \, {\vert} F_{1} {\vert}^{2}
     \nonumber \\ & + &
       ( p_{p} \, {\cdot} \, p_{\Lambda} + m_{p} \, m_{\Lambda} ) \, {\vert} G_{1} {\vert}^{2}
     \nonumber \\ & - &
         m_{p} \, k_{W} \, ( F_{1} \, G_{1}^{\dagger} + G_{1} \, F_{1}^{\dagger} )
     \label{htensor-pp},
     \end{eqnarray}
     %%%%% ------------------------------------
     %%%%% ------------------------------------
     \begin{eqnarray}
     H_{-,-} & = &
       ( p_{p} \, {\cdot} \, p_{\Lambda} - m_{p} \, m_{\Lambda} ) \, {\vert} F_{1} {\vert}^{2}
     \nonumber \\ & + &
       ( p_{p} \, {\cdot} \, p_{\Lambda} + m_{p} \, m_{\Lambda} ) \, {\vert} G_{1} {\vert}^{2}
     \nonumber \\ & + &
         m_{p} \, k_{W} \, ( F_{1} \, G_{1}^{\dagger} + G_{1} \, F_{1}^{\dagger} )
     \label{htensor-mm},
     \end{eqnarray}
     %%%%% ------------------------------------
     %%%%% ------------------------------------
     \begin{eqnarray}
     H_{t,t} & = &
       ( p_{p} \, {\cdot} \, p_{\Lambda} + m_{p} \, m_{\Lambda} ) \,
       \frac{ m_{p}^{2} }{ q^{2} } \, \big\{
      {\vert} F_{23} {\vert}^{2}
     - \frac{ q^{2} }{ m_{p}^{2} } \,  {\vert} G_{1} {\vert}^{2}
     \nonumber \\ & - &
        \frac{ m_{\Lambda} - m_{p} }{ m_{p} } \,
        \big[ \, F_{1} \, F_{23}^{\dagger} + F_{1}^{\dagger} \, F_{23}
      +  \frac{ E_{W} }{ m_{p} } \, ( {\vert} F_{1} {\vert}^{2}
      + {\vert} G_{1} {\vert}^{2} ) \, \big] \, \big\}
     \nonumber \\ & + &
       ( p_{p} \, {\cdot} \, p_{\Lambda} - m_{p} \, m_{\Lambda} ) \,
       \frac{ m_{p}^{2} }{ q^{2} } \, \big\{
      {\vert} G_{23} {\vert}^{2}
     - \frac{ q^{2} }{ m_{p}^{2} } \,  {\vert} F_{1} {\vert}^{2}
     \nonumber \\ & - &
        \frac{ m_{\Lambda} + m_{p} }{ m_{p} } \,
        \big[ \, G_{1} \, G_{23}^{\dagger} + G_{1}^{\dagger} \, G_{23}
     -  \frac{ E_{W} }{ m_{p} } \,( {\vert} F_{1} {\vert}^{2}
     + {\vert} G_{1} {\vert}^{2} ) \, \big] \, \big\}
     \label{htensor-tt},
     \end{eqnarray}
     %%%%% ------------------------------------
     %%%%% ------------------------------------
     \begin{eqnarray}
     H_{t,0} & = &
        \frac{ m_{p} \, k_{W} }{ q^{2} } \,  \big\{
        ( m_{\Lambda}^{2} - m_{p}^{2} ) \,
        ( {\vert} F_{1} {\vert}^{2} + {\vert} G_{1} {\vert}^{2} )
     %%%%% ------------------------------------
     \nonumber \\ & + &
           ( p_{p} \, {\cdot} \, p_{\Lambda} + m_{p} \, m_{\Lambda} ) \, \big[
           \frac{ m_{\Lambda} - m_{p} }{ m_{p} } \, F_{1} - F_{23} \big] \, F_{2}^{\dagger}
     \nonumber \\ & + &
           ( p_{p} \, {\cdot} \, p_{\Lambda} - m_{p} \, m_{\Lambda} ) \, \big[
           \frac{ m_{\Lambda} + m_{p} }{ m_{p} } \, G_{1} - G_{23} \big] \, G_{2}^{\dagger}
     \nonumber \\ & - & m_{p}^{2} \, \big(
           \frac{ m_{\Lambda} + m_{p} }{ m_{p} } \, F_{23} \, F_{1}^{\dagger}
         + \frac{ m_{\Lambda} - m_{p} }{ m_{p} } \, G_{23} \, G_{1}^{\dagger} \big) \, \big\}
     \label{htensor-tz},
     \end{eqnarray}
     %%%%% ------------------------------------
     %%%%% ------------------------------------
     \begin{eqnarray}
     H_{0,t} & = &
      H_{t,0} \, ( F_{i} \, {\leftrightarrow} \, F_{i}^{\dagger} , \,
                   G_{i} \, {\leftrightarrow} \, G_{i}^{\dagger} )
     \label{htensor-zt},
     \end{eqnarray}
     %%%%% ------------------------------------
     %%%%% ------------------------------------
     \begin{eqnarray}
     H_{0,0} & = &
          \frac{ m_{p}^{2} }{ q^{2} } \, k_{W}^{2} \,
           \big\{ 2 \, {\vert} F_{1} {\vert}^{2} \,
       +   \frac{ m_{\Lambda} + m_{p} }{ m_{p} } \,
           ( F_{1} \ F_{2}^{\dagger} + F_{2} \ F_{1}^{\dagger} ) \big\}
     \nonumber \\ & + &
           \frac{ m_{p}^{2} }{ q^{2} } \, k_{W}^{2} \,
           \big\{ 2 \, {\vert} G_{1} {\vert}^{2}
       +   \frac{ m_{\Lambda} - m_{p} }{ m_{p} } \,
          ( G_{1} \, G_{2}^{\dagger} + G_{2} \, G_{1}^{\dagger} ) \big\}
     \nonumber \\ & + &
          ( p_{p} \, {\cdot} \, p_{\Lambda} + m_{p} \, m_{\Lambda} ) \,
          \big\{ {\vert} G_{1} {\vert}^{2}
       +  \frac{ k_{W}^{2} }{ q^{2} } \, {\vert} F_{2} {\vert}^{2} \big\}
     \nonumber \\ & + &
         ( p_{p} \, {\cdot} \, p_{\Lambda} - m_{p} \, m_{\Lambda} ) \,
         \big\{ {\vert} F_{1} {\vert}^{2}
       + \frac{ k_{W}^{2} }{ q^{2} } \, {\vert} G_{2} {\vert}^{2} \big\}
     \label{htensor-zz},
     \end{eqnarray}
     %%%%% ------------------------------------
     %%%%% ------------------------------------
     \begin{eqnarray}
     F_{23} & = &
           \frac{ E_{W} }{ m_{p} }     \, F_{2}
         + \frac{ q^{2} }{ m_{p}^{2} } \, F_{3}
     \label{combination-f23}, \\
     G_{23} & = &
           \frac{ E_{W} }{ m_{p} }     \, G_{2}
         + \frac{ q^{2} }{ m_{p}^{2} } \, G_{3}
     \label{combination-f23},
     \end{eqnarray}
     %%%%% ------------------------------------
     %%%%% ------------------------------------
     \begin{eqnarray}
     2 \, ( p_{p} \, {\cdot} \, p_{\Lambda} + m_{p} \, m_{\Lambda} )
     & = &
     ( m_{p} + m_{\Lambda} )^{2} - q^{2}
     \label{m1.m2+p1.p2}, \\
     2 \, ( p_{p} \, {\cdot} \, p_{\Lambda} - m_{p} \, m_{\Lambda} )
     & = &
     ( m_{\Lambda} - m_{p} )^{2} - q^{2}
     \label{m1.m2-p1.p2}, \\
     2 \, m_{p} \, E_{W}
     & = &
     m_{p}^{2} - m_{\Lambda}^{2} + q^{2}
     \label{ew}, \\
     4 \, m_{p}^{2} \, k_{W}^{2}
     & = &
    {\lambda}(m_{p}^{2}, \, m_{\Lambda}^{2}, \, q^{2})
     \label{kw}.
     \end{eqnarray}
     %%%%% ------------------------------------

\begin{figure}[h]
     \includegraphics[width=0.4\textwidth]{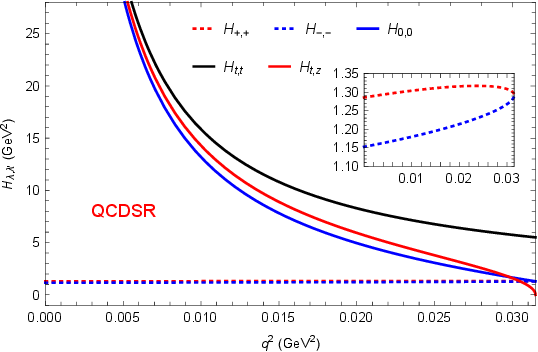} \quad
     \includegraphics[width=0.4\textwidth]{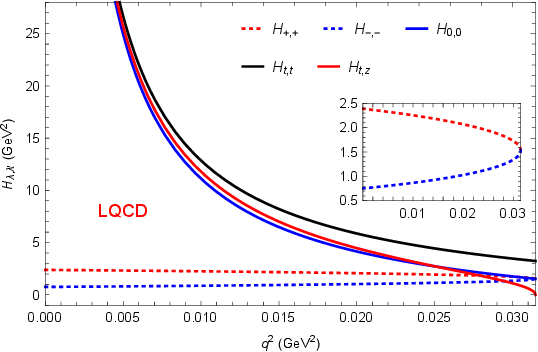}
     \caption{The distribution of $H_{{\lambda},{\lambda}^{\prime}}$
     versus $q^{2}$,
     where the lines in left (right) figure are calculated
     with the center values of the QCDSR (LQCD) form
     factors of Ref. \cite{PRD.112.094035}
     (Ref. \cite{PRL.135.231901}).}
     \label{fig:hijq2}
     \end{figure}

     The distribution of amplitude tensors
     $H_{{\lambda},{\lambda}^{\prime}}$
     for the $p$ ${\to}$ ${\Lambda}$ transition
     versus $q^{2}$ is illustrated in Fig. \ref{fig:hijq2}.
     %%%%% ------------------------------------
     %%%%% ------------------------------------

      \end{appendix}

     %------------------------------------
     
     %------------------------------------

     \end{document}